\documentclass[prl,superscriptaddress,reprint,amsmath,amssymb,aps,longbibliography]{revtex4-1}

\usepackage{xcolor}
\usepackage{ulem}
\usepackage{graphicx}
\usepackage{float}
\usepackage{diagbox}
\usepackage[
pdfstartview=FitH,
bookmarks=true,
breaklinks=true,
colorlinks=true,
linkcolor=blue,anchorcolor=blue,
citecolor=blue,filecolor=blue,
menucolor=blue,urlcolor=blue]{hyperref}

\begin{document}

\title{Observation of universal dissipative dynamics in strongly correlated quantum gas}

\author{Yajuan Zhao}
\thanks{These authors contribute equally to this work.}
\author{Ye Tian}
\thanks{These authors contribute equally to this work.}
\author{Jilai Ye}
\thanks{These authors contribute equally to this work.}
\affiliation{Department of Physics and State Key Laboratory of Low Dimensional Quantum Physics, Tsinghua University, Beijing, 100084, China}
\author{Yue Wu}
\affiliation{Institute for Advanced Study, Tsinghua University, Beijing, 100084, China}
\author{Zihan Zhao}
\author{Zhihao Chi}
\author{Tian Tian}
\affiliation{Department of Physics and State Key Laboratory of Low Dimensional Quantum Physics, Tsinghua University, Beijing, 100084, China}
\author{Hepeng Yao}
\affiliation{DQMP, University of Geneva, 24 Quai Ernest-Ansermet, CH-1211 Geneva, Switzerland}
\author{Jiazhong Hu}
\email{hujiazhong01@ultracold.cn}
\affiliation{Department of Physics and State Key Laboratory of Low Dimensional Quantum Physics, Tsinghua University, Beijing, 100084, China}
\affiliation{Frontier Science Center for Quantum Information and Collaborative Innovation Center of Quantum Matter, Beijing, 100084, China}
\author{Yu Chen}
\email{ychen@gscaep.ac.cn}
\affiliation{Graduate School of China Academy of Engineering Physics, Beijing, 100193, China}
\author{Wenlan Chen}
\email{cwlaser@ultracold.cn}
\affiliation{Department of Physics and State Key Laboratory of Low Dimensional Quantum Physics, Tsinghua University, Beijing, 100084, China}
\affiliation{Frontier Science Center for Quantum Information and Collaborative Innovation Center of Quantum Matter, Beijing, 100084, China}

\date{\today}

\begin{abstract}

Dissipation is unavoidable in quantum systems. It usually induces decoherences and changes quantum correlations. To access the information of strongly correlated quantum matters, one has to overcome or suppress dissipation to extract out the underlying quantum phenomena. However, here we find an opposite effect that dissipation can be utilized as a powerful tool to probe the intrinsic correlations of quantum many-body systems. Applying highly-controllable dissipation in ultracold atomic systems, we observe a universal dissipative dynamics in strongly correlated one-dimensional quantum gases. The total particle number of this system follows a universal stretched-exponential decay, and the stretched exponent measures the anomalous dimension of the spectral function, a critical exponent characterizing strong quantum fluctuations of this system. 
This method could have broad applications in detecting strongly correlated features, including spin-charge separations and Fermi arcs in quantum materials.

\end{abstract}

\maketitle

Open quantum systems and non-Hermitian physics are emerging topics in recent years, uncovering fascinating dissipation-driven phenomena and attracting considerable attention \cite{Diehl2008,RevModPhys.88.021002,ashida2020non,Bergholtz2021RMP}. Ultracold atomic gases are among the best testbeds for exploring dissipation effects in quantum many-body systems because of the full control of both dissipation channels and their strengths, together with the versatile tunability of many-body quantum correlations. Utilizing these advantages, novel physics different from those in closed systems have been observed in cold atom platforms, including dissipative-stabilized strong-correlated matters \cite{Leonard2017,Ma2019,Ferioli2023,dogra2023universal}, quantum phase transitions in superradiance \cite{Hall1999,Baumann2010,Zhang2021,Konishi2021,Lu2023}, continuous time crystals \cite{PhysRevLett.127.043602,Kongkhambut2022}, and dissipation-enabled entangled states \cite{Lin2013,PhysRevLett.107.080503}. However, these novel physics usually emerge in the strong dissipation regimes and dissipation-driven steady states. In contrast, in the weak dissipation regime and the short-time scale before reaching the steady state, how dissipation interplays with quantum many-body correlations is rarely studied.

A weak dissipation usually does not disturb intrinsic quantum correlations at short time. However, it does cause a dynamical response, usually manifested as the decay of a physical observable in time. With constant dissipation, we often observe exponential decay  for most conventional quantum states. These states are quantum phases with well-defined quasi-particles, displaying delta-function type single-particle spectral function. In contrast, if quantum fluctuation in the system is so strong that it destroys well-defined quasi-particles, the dissipative dynamics is predicted to display a stretched-exponential decay \cite{pan2020non}. Especially for quantum critical states \cite{Sachdev2011,giamarchi2003quantum,JiangYu-Zhu,LIANG20222550}, the spectral function exhibits a power-law divergence around the threshold $\Delta$ in the form of $(\omega^2-\Delta^2)^{-\eta}$, and the dissipative dynamics at short time follows a stretched-exponential form of $\exp[-(t/\tau_0)^{2\eta-1}]$ \cite{pan2020non}. The power $\eta$ in the spectral function is the anomalous dimension, a critical exponent manifesting strong quantum fluctuations in this system. Usually, the stronger the quantum fluctuation, the smaller the anomalous dimension $\eta$. Thus, measuring the dynamics under weak dissipation provides an alternating route to access many-body correlations at equilibrium, in contrast to existing measurements in condensed matter and cold atom physics using Hermitian perturbation as a probing tool.

The one-dimensional (1D) Bose gas is an optimal choice for performing measurements of strong correlation effects by using dissipation. First, 1D systems are known to display strong quantum fluctuations at low temperature and obey the divergent single-particle spectral function mentioned above \cite{giamarchi2003quantum,JiangYu-Zhu}. Moreover, its anomalous dimension is tunable by tuning interaction strength \cite{giamarchi2003quantum}. Secondly, the 1D Bose gas is integrable, and the analytical solution allows a quantitative comparison between theory and experiment. This advantage allows us to benchmark this new measurement scheme. Thirdly, 1D atomic gas can be precisely controlled, and many fascinating phenomena have been observed already \cite{kinoshita2004observation,paredes2004tonks,Haller2009,Haller2010,Meinert2017,yang2017quantum,Erne2018,Wilson2020,Vijayan2020,senaratne2022spin,SeeToh2022,le2022direct,Cavazos2023,1d-thermometer,yao-crossoverD-2022,PhysRevLett.122.170403}. However, all existing experiments focus on density distributions, momentum distributions, or density and spin correlation functions. The measurement of single-particle spectral function is still missing. Especially, the crucial quantity of the anomalous dimension has not been experimentally measured.

\begin{figure*}[tb]
    \includegraphics[width=1\textwidth]{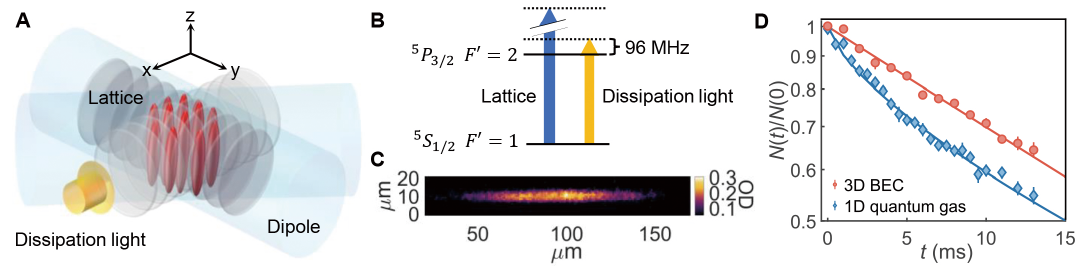}
    \centering
    \caption{\textbf{Illustration of the experiment setup.} (\textbf{A}), A BEC with $4\times 10^4$ $^{87}$Rb atoms is prepared in a crossed dipole trap and loaded into 2D optical lattice made from two orthogonal laser beams. 3300 tubes of 1D Bose gas ensemble are created. A near-resonant light along the $x$ axis is shined onto the ensemble to introduce one-body dissipation. An absorption imaging is set up along the $x$ axis. (\textbf{B}), The dissipation light is set to 96 MHz blue-detuned from $|5S_{1/2},F=1\rangle$ $\rightarrow$ $|5P_{3/2},F=2\rangle$ transition. 
    (\textbf{C}), A typical time-of-flight (TOF) image for atoms in 2D array of 1D tubes.
    (\textbf{D}), Normalized atom number $N(t)/N(0)$ versus dissipation-duration time $t$ in log-linear plot for the dissipative dynamics of 1D quantum gas (blue diamonds) and 3D BEC (red circles). The solid curves are fittings using the stretched-exponential function, with the fitted stretched exponents $\alpha=0.70(4)$ (blue) and $\alpha=0.99(4)$ (red). 
    This shows qualitative difference of the stretched-exponential decay versus the exponential one between strongly correlated one-dimensional gas and weakly interacting three-dimensional BEC.
    All error bars correspond to one standard deviation.}
    \label{setup}
\end{figure*}

In our experiment, we realize Luttinger liquid of bosonic $^{87}$Rb atoms trapped in a 1D array of tubes created by a two-dimensional (2D) optical lattice (see Supplementary materials (SM), and Fig.~\ref{setup}\textbf{A}). These atoms, initialized in the state of $|5S_{1/2},F=1,m_F=1\rangle$, are confined in a crossed dipole trap and levitated against gravity using a magnetic field gradient. By adjusting the intensity of the lattice beams, we tune the lattice depth $U_{l}$ of each lattice beam to modify the transverse ground state size $a_{\bot}$. This leads to the change of the 1D interaction strength $g_\textrm{1D}=4\hbar^2 a_s/[m a^2_{\bot}(1-1.46\frac{a_s}{a_{\bot}})]$, where $m$ is the mass of an atom, and $a_s$ is the three-dimensional scattering length of $^{87}$Rb at 98 Bohr radii. Thus, as we change the lattice depth $U_{l}$ from $0.52$~kHz to $430$~kHz, the Luttinger liquid is tuned from weakly interacting region to Tonks-Girardeau region. The mean dimensionless interaction strength $\gamma=mg_{\textrm{1D}}/n\hbar^2$, representing the ratio of interaction energy and kinetic energy, is varied from $0.08$ to $5.36$, where $n$ is the 1D atomic density. To introduce well-controlled one-body loss and avoid two-body loss caused by light-assisted collision, we set the frequency of the dissipation light to be 96~MHz blue-detuned from the transition of $|5S_{1/2},F=1\rangle$$\rightarrow$$|5P_{3/2},F=2\rangle$ (see Fig.~\ref{setup}\textbf{B}).

The loss introduced by such dissipation light leads to decay of the atom number in the Luttinger liquid. We fix the intensity of the dissipation light, and measure the residual atom number $N(t)$ with respect to the duration time $t$ of the dissipation light. As shown in Fig.~\ref{setup}\textbf{D}, a decay process (blue diamonds) different from the conventional exponential decay is observed. Here, we set the lattice depth $U_l$ to $65$~kHz with a corresponding $\gamma$ of $1.54$, and fix the intensity of the dissipation light to a saturation parameter $s$ of $0.01$. In order to investigate this unconventional decay behavior, we fit the decay curve to both stretched-exponential function and power function, and find the stretched-exponential function an obviously better fit. Inspired by Ref.~\cite{pan2020non}, we use the fitting function
\begin{equation}
    N(t)=N(0)\ {\textrm{exp}}\left[-\left(\frac{t}{\tau_0}\right)^\alpha\right], \label{eq:stretch}
\end{equation}
to extract stretched exponent $\alpha$, dissipation characteristic time $\tau_{0}$, and initial atom number $N(0)$. $\alpha=0.70\pm0.04$ and $\tau_0=25.3\pm1.1$~ms are obtained from fitting.
For comparison, we also measure the loss dynamics of a three-dimensional Bose-Einstein condensate (BEC) in a crossed dipole trap where the decay follows an exponential function with the fitted stretched exponent $\alpha=0.99\pm0.04$ (red circles in Fig.~\ref{setup}\textbf{D}).

\begin{figure*}[htbp]
    \centering
    \includegraphics[width=1\textwidth]{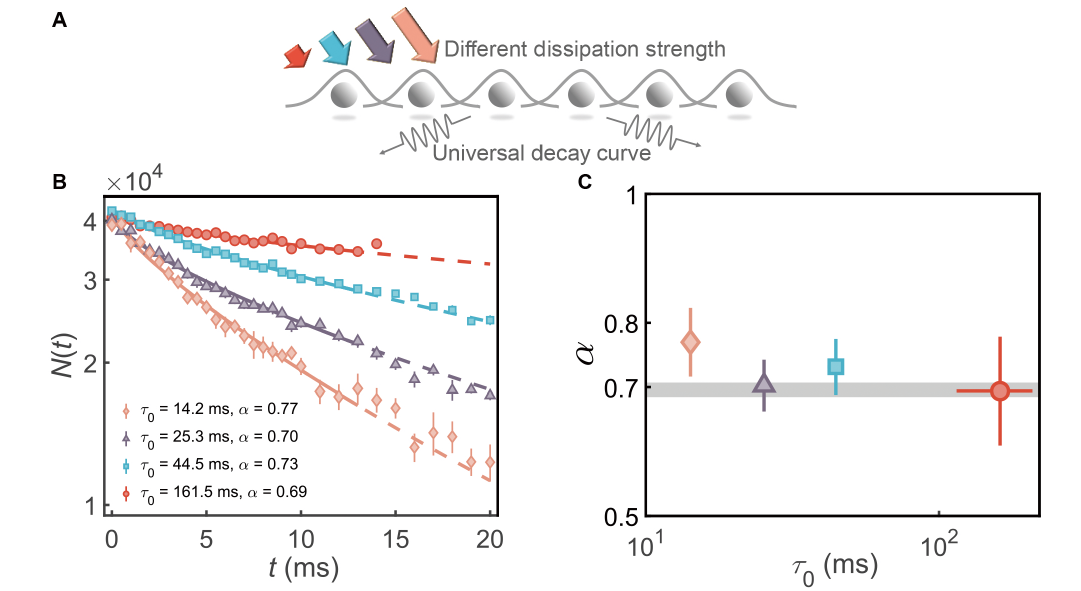}
    \caption{\textbf{Dissipation dynamics at different dissipation strength for Luttinger liquid at dimensionless interaction strength $\gamma=1.54$.} (\textbf{A}), Illustration of the dissipation process at different dissipation strength. Arrows with different colors imply different dissipation strengths applied to the 1D Bose gas. (\textbf{B}), The atom number $N(t)$ decays with respect to the dissipation time $t$ in log-linear plot. Data with different labels represent dissipation processes at different dissipation strengths. The solid lines are fittings to the stretched-exponential decay in Eq.~\ref{eq:stretch} using data within the upper bound of the dissipation time $t_{u}=13$~ms. (\textbf{C}), The fitted $\alpha$ under different dissipation strength characterized by $\tau_0$. Vertical and horizontal error bars are the one standard deviations of the fitted $\alpha$-s and $\tau_0$-s, respectively. The grey shade is the theoretical prediction of the non-Hermitian linear response theory, taking both finite-time correction and the atom-number inhomogeneity of different tubes into account. All error bars correspond to one standard deviation.
    }
    \label{differentratio}
\end{figure*}

Then, we change the dissipation rate in the Luttinger liquid by tuning the intensity of the dissipation light, so that the dissipation characteristic time $\tau_{0}$ is ranged from $10$~ms to $200$~ms (Fig.~\ref{differentratio}\textbf{A} and \textbf{B}). At different dissipation rates, the atom number decay fittings show consistent stretched exponent $\alpha$-s, remaining the same within one standard deviation (Fig.~\ref{differentratio}\textbf{C}). These data demonstrate that the stretched-exponential decay is a universal response of the Luttinger liquid to the dissipative probe. The stretched exponent $\alpha$ being independent of the dissipation strength, indicates that this exponent measures an intrinsic equilibrium property of the Luttinger liquid.

\begin{figure*}[htbp]
    \centering
    \includegraphics[width=1\textwidth]{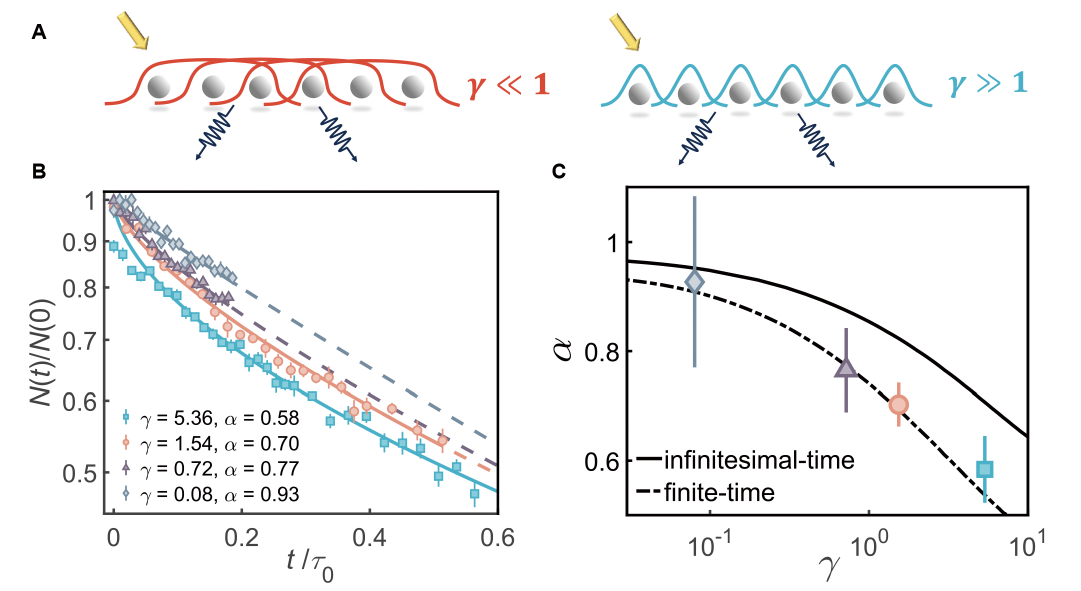}
    \caption{\textbf{Dissipation dynamics at different dimensionless interaction strength $\gamma$.} (\textbf{A}), Illustration of the dissipation process at different dimensionless interaction strength $\gamma$. (\textbf{B}), The normalized atom number $N(t)/N(0)$ versus the rescaled time $t/\tau_0$ at different $\gamma$ in log-linear plot. Different labels show dissipation processes under different $\gamma$ and the solid lines are fits to Eq.~\ref{eq:stretch} using data within the upper bound of the dissipation time $t_{u}$. 
    (\textbf{C}), The fitted stretched exponent $\alpha$ as a function of $\gamma$ in linear-log plot. The black solid line shows the zero-temperature theoretical prediction for $\alpha=2\eta-1$ at different $\gamma$, while the dashed line shows the prediction with the finite-time correction. The grey diamonds with $\gamma=0.08$ are in the crossover of the three-dimensional and the one-dimensional Bose gas region, and the theoretical prediction based on Lieb-Liniger model is not applicable to these data. All error bars correspond to one standard deviation.} 
    \label{diffgamma}
\end{figure*}

In order to explore the relation between the dissipative dynamics and the intrinsic correlations of the Luttinger liquid, we investigate how the stretched exponent $\alpha$ changes in different interaction regimes, tuning the Luttinger liquid from weakly interacting Bose gas to strongly correlated Tonks-Girardeau gas (Fig.~\ref{diffgamma}\textbf{A}). Here, Luttinger liquid with the dimensionless interaction strength $\gamma$ tuned from $0.08$ to $5.36$ are created, where we measure the atom decay curves (Fig.~\ref{diffgamma}\textbf{B}), fit the data with Eq.~\ref{eq:stretch}, and extract the stretched exponent $\alpha$ from fittings at these dimensionless interaction strengths, as shown in Fig.~\ref{diffgamma}\textbf{C}. 
Here, we obtain $\alpha=0.93\pm0.16$, $0.77\pm0.08$, $0.70\pm0.04$, and $0.58\pm0.06$ for $\gamma$ = 0.08, 0.72, 1.54, and 5.36, respectively.

These observations can be quantitatively understood by applying the non-Hermitian linear response theory to the Luttinger liquid \cite{pan2020non, JiangYu-Zhu}. The theory connects dissipative dynamics of an observable to the correlation functions of the initial equilibrium state. By calculating the linear-order perturbations and spectral functions at zero temperature (see SM), the theoretical model predicts a stretched-exponential behavior for one-body dissipation as $N(t)\propto \textrm{exp}[-(t/\tau_0)^{2\eta-1}]$. 
This prediction connects the stretched exponent $\alpha$ with the anomalous dimension $\eta$ as $\alpha=2\eta-1$, where $\eta$ is determined by Luttinger parameter $K$ as $\eta=1-1/\left(4K\right)$. Using Lieb-Liniger model, both $\eta$ and $K$ are determined by the dimensionless interaction strength $\gamma$ (see SM).
%, where $\eta$ is only decided by $\gamma$ and characterizes the asymmetry of the spectral function $A(\omega)$ $\propto \Theta(\omega-\omega_0)/(\omega-\omega_0)^\eta$ of 1D Luttinger liquid with step function $\Theta(\omega)$. 
%Of course, the spectral function at finite temperature is modified at low frequencies, while the short time dynamics is attributed relative high frequency spectrum which is not altered. We find the concerned dynamics is not influenced when $t<\beta/(1-\eta)$, which is typically 10 $ms$. Meanwhile the extremal short-time scale dynamics is determined by high frequency spectral function, where linear Luttinger liquid approximation fails to be exact.
In Fig.~\ref{diffgamma}\textbf{C}, we plot the theoretical predicted values for zero-temperature case of $\alpha=2\eta-1$ versus $\gamma$ (black solid curve). As the dimensionless interaction strength $\gamma$ increases, the quantum fluctuations get stronger, and the quasi-particle description fails. This leads to a decreasing anomalous dimension $\eta$ with a stronger deviation from 1, and a smaller Luttinger parameter $K$.
This zero-temperature theoretical prediction curve is not fully aligned with experimental data due to the finite-time effect, because it is only valid at dissipation time $t\ll h/\pi k_B T$, where $h/\pi k_B T \sim 3$~ms for a typical temperature $T=5$~nK (see SM). 
%\chen{(Due to the technical limit, / As the extremal short-time scale dynamics is determined by high frequency spectral function, where linear Luttinger liquid approximation we use fails to be exact,)}
However, to guarantee reasonable data quality, experimentally we have to choose an intermediate dissipation time $t\sim h/\pi k_B T$, and the fitted stretched exponent $\alpha$ is influenced by the chosen upper bound $t_u$ of the dissipation time (see Fig.~S2\textbf{B}). Therefore, we extend the zero-temperature linear response theory into the finite-temperature regime with higher-order corrections, and plot the corrected theoretical curve of $\alpha$ in Fig.~\ref{diffgamma}\textbf{C} (black dashed curve) and Fig.~\ref{differentratio}\textbf{C} (grey shade). For the fittings to extract $\alpha$, we only include the experimental data within $t_{u}$, represented by the solid lines in both Fig.~\ref{differentratio}\textbf{B} and \ref{diffgamma}\textbf{B}. The corrected theoretical predictions agree well with our experimental data. One important thing to be noticed is that this finite-time correction is temperature-independent. We choose the interrogate time upper bound $t_u$ to be inversely proportional to the temperature $T$ of the system to keep the product of $t_u$ and $T$ as a constant (see SM). 
%\zhao{To be mentioned, for the case (grey diamonds in Fig.~\ref{diffgamma}\textbf{b} and \ref{diffgamma}\textbf{c}) where the lattice depth is too small so that the tunneling between different tubes is non-negligible, the theoretical prediction is invalid because the system is no longer in the 1D region. Temperature calibration of 1D Bose gas (see Methods) also becomes invalid in this case, so we can not accurately measure its temperature and choose the corresponding time upper bound $t_u$. Actually, we use experimental data within a mediate $t_u=10$~ms for the fitting to extract $\alpha$ in this weak-interaction regime.}

\begin{figure*}[htbp]
\centering
    \includegraphics[width=1\textwidth]{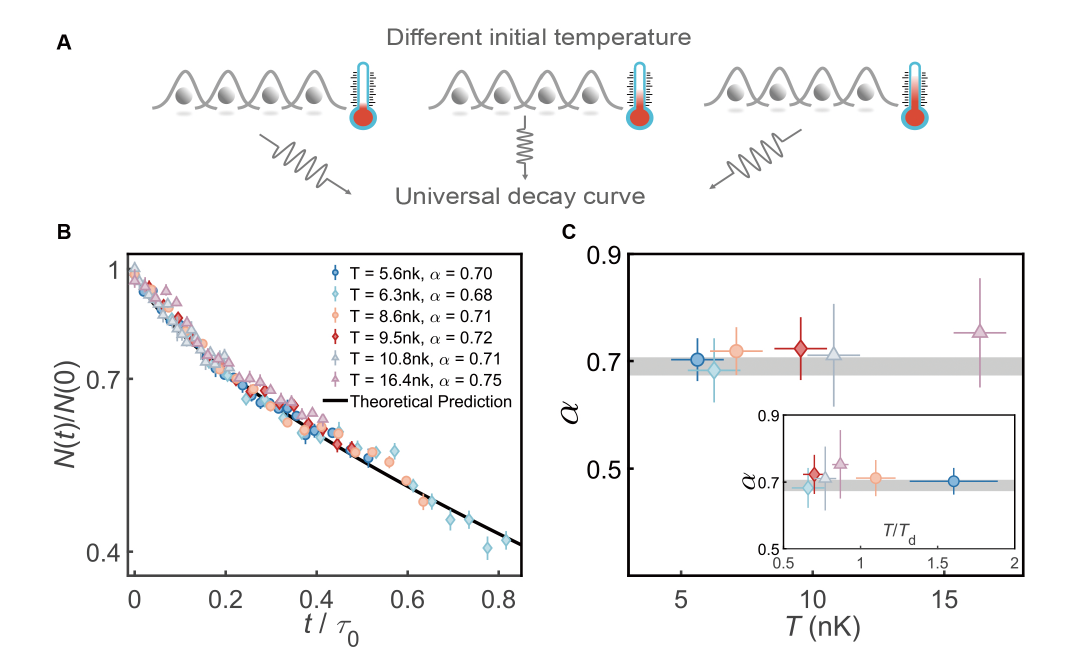}
    \caption{\textbf{Dissipation dynamics at different temperature.}
    (\textbf{A}), Illustration of the dissipation processes at different temperature $T$.
    (\textbf{B}), The normalized atom number $N(t)/N(0)$ versus the rescaled dissipation time $t/\tau_0$ in log-linear plot at different temperature. All data fall on the same stretched-exponential curve with $\alpha=0.69$. All error bars correspond to one standard deviation.
    (\textbf{C}), Fitted $\alpha$ at different temperatures. The horizontal error bars stand for the uncertainty of the temperature, while the vertical error bars represent the one-standard-deviation confidence intervals of the fittings. The grey shade is estimated by the non-Hermitian linear response theory with the finite-time correction and the atom-number inhomogeneity. The inset shows the fitted $\alpha$ versus the rescaled temperature $T/T_{d}$, where $T_d = \hbar^2n^2/2mk_B$.
    }
    \label{diffrenttemperature}
\end{figure*}

To further demonstrate the universal dissipation dynamics is robust against the temperature variation and thermal effects, we perform the measurements for Luttinger Liquid with the same $\gamma$ but different temperature $T$ (Fig.~\ref{diffrenttemperature}\textbf{A}). To achieve this, we vary the depth of the crossed dipole trap while simultaneously adjust the depth of the lattice, to keep the dimensionless interaction strength $\gamma$ constant at $1.63 \pm 0.10$. 
As a result, we are able to tune the initial temperature of the ensemble from 5.6~nK to 16.4~nK (see SM for temperature calibration), and show consistent $\alpha$ values for different temperatures (Fig.~\ref{diffrenttemperature}\textbf{C}). 
We also rescale the dissipation time with the dissipation characteristic time $\tau_{0}$ for data at each temperature, normalize the atom number, and find that the different temperature groups fall on the same line with $\alpha=0.69$, which agrees well with the theoretical prediction of $\alpha$ at $\gamma=1.63$ (Fig.~\ref{diffrenttemperature}\textbf{B}). This observation demonstrates the universality of the dissipation process among a wide range of temperature. For each parameter set, we calculate the chemical potential of 1D tubes using the Yang-Yang equation \cite{yang2017quantum}, confirming that the chemical potential and the temperature are small enough compared with the tightly confined vibrational frequency and thus all ensembles can be described by one-dimensional Luttinger liquid.

In addition, we normalize the temperature $T$ with the degenerate temperature $T_d = \hbar^2n^2/2mk_B$, and plot $\alpha$ versus this normalized temperature in Fig.~\ref{diffrenttemperature}\textbf{C} inset. According to the theoretical prediction of the Lieb-Liniger model at finite temperature \cite{cheng2022one,Cheng2022Exact}, the normalized temperature $T/T_d$ characterizes the extent of deformation in the spectral function caused by thermal effects. When $T/T_d$ is smaller than $1.5$, the spectral function remains almost unchanged compared with the zero-temperature case. Notably, our finite temperature results agree with this theoretical prediction, since the stretched exponent $\alpha$, characterizing the spectral function, remains constant when $T/T_d$ falls within the range of $0.6$ to $1.6$.

Our experiment demonstrates a universal dissipative dynamics in strongly correlated quantum gas, while the disspative process is only decided by the intrinsic correlations, independent of any external parameters such as dissipation strength and temperature.
Similar phenomenon named anomalous decay of coherence was also observed in the superfluid-Mott insulator phase transition \cite{Bouganne2020}. Such anomalous decay was attributed to the diffusion dynamics in momentum space in the original paper, and later explained by 
the non-Hermitian linear response theory \cite{pan2020non}. In this article, we fully benchmark the unconventional decay behavior in 1D Bose gas experimentally, and show that the predictions of the non-Hermitian linear response theory agree well quantitatively with the experimental results. These results fully support the validity of this new theory, and extend this theory from zero temperature region to finite temperature region.

To conclude, we demonstrate a powerful method to utilize dissipation to probe quantum systems and observe equilibrium many-body quantum correlations. This novel dissipative probe method is used to detect the anomalous dimension in 1D Bose gas experimentally, which is hardly accessible in a closed system due to technical difficulties \cite{1d-thermometer}. This method is also expected to detect other quantum correlation features in strongly correlated systems with fractional excitations, including anomalous dimensions in one-dimensional Hubbard model with spin-charge separation and Fermi arc in high-TC superconductors.

\noindent{\textbf{Acknowledgement}}
We acknowledge the inspiring discussions with H. Zhai, X. -W. Guan, and T. Giamarchi, and the technical supports from W. Zhang and Z. Zhang.\\
\noindent{\textbf{Funding}}:
This work is supported by National Natural Science Foundation of China (92165203, 61975092, 11974202, 12174358),
 National Key Research and Development Program of China (2021YFA1400904, 2021YFA0718303, 2022YFA1405300),
 Beijing Natural Science Foundation (Z180013) and
 Swiss National Science Foundation (200020-188687).\\

\bibliography{ref.bib}

\end{document}

\author
{Yajuan Zhao,$^{1}\dagger$ Ye Tian,$^{1}\dagger$ Jilai Ye,$^{1}\dagger$ Yue Wu,$^{2}$  Zihan Zhao,$^{1}$ Zhihao Chi,$^{1}$\\ Tian Tian,$^{1}$ Hepeng Yao,$^{3}$ Jiazhong Hu,$^{1,4\ast}$ Yu Chen,$^{5\ast}$ Wenlan Chen,$^{1,4\ast}$
\\
\normalsize{$^{1}$Department of Physics and State Key Laboratory of Low Dimensional Quantum Physics},\\
\normalsize{Tsinghua University, Beijing, 100084, China}\\
\normalsize{$^{2}$Institute for Advanced Study, Tsinghua University, Beijing, 100084, China}\\
\normalsize{$^{3}$DQMP, University of Geneva, 24 Quai Ernest-Ansermet, CH-1211 Geneva, Switzerland}\\
\normalsize{$^{4}$Frontier Science Center for Quantum Information and Collaborative Innovation }\\
\normalsize{Center of Quantum Matter, Beijing, 100084, China}\\
\normalsize{$^{5}$Graduate School of China Academy of Engineering Physics, Beijing, 100193, China}\\
\\
\normalsize{$\dagger$These authors contribute equally to this work.}\\
\normalsize{$^\ast$To whom correspondence should be addressed; E-mail: hujiazhong01@ultracold.cn, }\\
\normalsize{ychen@gscaep.ac.cn, cwlaser@ultracold.cn.}
}
\baselineskip24pt


\maketitle


% In setting up this template for *Science* papers, we've used both
% the \section* command and the \paragraph* command for topical
% divisions.  Which you use will of course depend on the type of paper
% you're writing.  Review Articles tend to have displayed headings, for
% which \section* is more appropriate; Research Articles, when they have
% formal topical divisions at all, tend to signal them with bold text
% that runs into the paragraph, for which \paragraph* is the right
% choice.  Either way, use the asterisk (*) modifier, as shown, to
\newpage
\section*{Materials and Methods}
\uline{Preparation of Luttinger liquid}

%\subsection*{Preparation of Luttinger liquid}
We start with a Bose-Einstein condensate (BEC) of $N=4\times 10^4$ $^{87}$Rb atoms at $|5S_{1/2},F=1,m_{F}=1\rangle$, confined in a horizontally crossed dipole trap with a wavelength at 1064~nm. The waist of each dipole beam is 50~$\mu\textrm{m}$ and the angle between them is $45^{\circ}$, which induces an anisotropic 3D harmonic trap. The trap depth is $U_{z}/h=6.8~\textrm{kHz}$ in total and its vibrational frequencies along three dimensions are $\omega_x=2\pi\times19.3~\textrm{Hz}$, $\omega_y=2\pi\times13.5~\textrm{Hz}$, and $\omega_z=2\pi\times33.0~\textrm{Hz}$ respectively. Meanwhile, we introduce a magnetic field gradient at 30.7~Gauss/cm to compensate for the gravity. The BEC is then transformed into 1D Luttinger liquid by adiabatically ramping up a 2D optical lattice with a wavelength of $776.5$~nm within 80~ms. These two lattice beams propagate perpendicularly to each other, with horizontal and vertical polarization respectively and frequency difference at $160$~MHz so that the interference between lattice beams can be avoided. The waist of the lattice beams is 400~$\mu\textrm{m}$, much larger than the atomic cloud size, to guarantee the homogeneity of radial vibrational frequency  $\omega_{\perp}$ among the ensemble. The 2D optical lattice, together with the crossed dipole trap, provides independent control of the radial and axial confinements of the 1D gas.
~\\\\
%\subsection*{Non-Hermitian linear response theory on Luttinger liquids}
\noindent
\uline{Non-Hermitian linear response theory on Luttinger liquids}

When a 1D system experiences a dissipative perturbation formed as $-i\kappa\hat{\mathcal O}^\dagger(z,t)\hat{\mathcal O}(z,t)$ starting from the initial time $t_i=0$ with $\kappa$ denoting the dissipation rate, any observable $\hat{\mathcal W}$ will have a measurable difference $\delta{\mathcal W}(t)$ at time $t$ compared with its value at the steady state without dissipation. $\delta{\mathcal W}(t)$ is expressed as 
\begin{equation}
%\begin{split}
\delta {\mathcal W}(t)=2\kappa\int dz\int_0^t d\tau\langle \hat{\mathcal O}^\dag(z,\tau)\hat{\mathcal W}(t)\hat{\mathcal O}(z,\tau) 
-\frac{1}{2}\{\hat{\mathcal O}(z,\tau)\hat{\mathcal O}(z,\tau),\hat{\mathcal W}(t)\}\rangle, \label{eq2}
%\end{split}
\end{equation}
where $\{,\}$ is the anti-commutator.
To describe one-body loss in our experiment, we use $\hat{\mathcal O}(z)=\int dk \hat{a}_k e^{ikz}$ where $\hat{a}_k$ is the bosonic annihilation operator at momentum mode $k$, and we choose the atom number $\hat{\mathcal W}=\hat{n}_k=\hat{a}^\dagger_k \hat{a}_k$ as the observable. Following the derivations in Ref~\cite{pan2020non}, Eq.~1 becomes
\begin{equation}
\delta n_k(t)=-2\kappa n_k(0)\int_0^t d\tau g_k(\tau)g_k(-\tau), \label{eq3}
\end{equation}
where $g_k(t)$ is the single-particle correlation function in the form of $g_k(t)=\langle [\hat{a}_k(t),\hat{a}^\dag_k(0)]\rangle$.
When the 1D system is described by the Luttinger liquid theory at zero temperature, the single-particle correlation function is dominant by the phase fluctuations and $g_k(t)$ has a form of $g_k(t)\sim t^{\eta-1}$, where $\eta$ is the anomalous dimension with $\eta=1-1/\left(4K\right)$. Here, $K$ is the Luttinger parameter and can be uniquely determined by the dimensionless interaction strength $\gamma$ (see supplementary text). Although $g_k(t)$ has weak dependence on momentum $k$, leading to $k$-dependent $\delta n_k(t)$, these $k$-dependent deviations from $t^{\eta-1}$ cancel each other when we integrate the whole momentum space. Consequently, the total atom number $N(t)$ decays in the form
\begin{eqnarray}
N(t)=N(0)\exp[-(t/\tau_0)^{2\eta-1}], \label{eq4}
\end{eqnarray}
where $\tau_0$ is a time constant.
%scale inversely proportional to $\kappa$.
When the dimensionless interaction strength $\gamma$ in the Luttinger liquid increases, the anomalous dimension $\eta$ decreases. This
leads to a smaller stretched exponent in the decay curve, deviating further from the conventional exponential decay. More detailed discussions and derivations can be found in the supplementary text.

It is worth noting that this theory has an upper limit and a lower limit for the time scale within which it could accurately describe the dissipation dynamics in 1D Bose gas. The upper bound of the timescale is on the order of tens of $1/\kappa$, and the extremely-short-timescale dynamics is determined by high-frequency spectral function, where linear Luttinger liquid approximation fails to be exact.
%The upper bound of the timescale is determined by the temperature and the anomalous dimension as $1/\left(\pi k_BT (1-\eta)\right)$, and the extremely-short-timescale dynamics is determined by high frequency spectral function, where linear Luttinger liquid approximation fails to be exact. 
Besides, the theoretical prediction fails when the dimensionless interaction strength becomes too small, because the system is no longer in the Tomonaga-Luttinger liquid (TLL) regime in this case.
~\\\\
%\subsection*{Finite-time corrections at finite temperature}
\noindent
\uline{Finite-time corrections at finite temperature}

When the system has a finite temperature $T$, the dynamics of the total atom number in Eq.~\ref{eq4} needs to be corrected into the following form (see supplementary text)
\begin{eqnarray}
  N(t)= N(0)\exp\bigg\{-\frac{1}{\tau_0^{2\eta-1}}\int_0^t{\textrm{d}}\tau\bigg[\frac{h}{\pi k_BT}  
\times\sinh\left(\frac{\pi k_BT\tau}{ h}\right)\bigg]^{2\eta-2}\bigg\}. \label{eq:finite-temperature}
\end{eqnarray}
Therefore the stretched-exponential form in Eq.~\ref{eq4} is only valid when $\pi k_BTt/h  \ll 1$. In our experiment, this requires $t \ll h/\pi k_BT \sim$ 3~ms for a typical temperature at $T=5$~nK. %As the extreme short-time-scale dynamics is determined by high-frequency spectral function where linear Luttinger-liquid theory fails, 
Due to the measurement noise,
experimentally we need to choose an intermediate dissipation duration time $t \sim h/\pi k_BT$, so a correction to the stretched exponent in the case of zero temperature is necessary. 
Practically, we set an upper bound $t_u$ of the dissipation time, to satisfy $\pi k_BTt_u/h = 5$. In other words, we keep $t_uT$ constant at $\mathcal{C}=5h/\pi k_B $ in the fittings of decay curves to extract the stretched exponent. The constant $\mathcal{C}$ is chosen to guarantee that the atom number difference between Eq.~\ref{eq4} and Eq.~\ref{eq:finite-temperature} is within $6\%$, comparable to the error bar of the measured atom number in our data. In Fig.~2\textbf{B} and Fig.~3\textbf{B}, the boundary between the solid lines and the dashed lines represents $t_u$. 
Then, by fitting the curve in Eq.~\ref{eq:finite-temperature} with a stretched-exponential function $ \exp\left[-(t/\tau_0)^{\alpha^*}\right]$ within $t_u$, we are able to extract a corrected theoretical prediction for the stretched exponent $\alpha^*$, which takes the finite-time effect into account, and is only determined by the dimensionless interaction strength $\gamma$ and the constant $\mathcal C$ $=t_u T$. The $\alpha^*$-s are shown in the black dashed line in Fig.~3\textbf{C}, and grey shades in Fig.~2\textbf{C} and Fig.~4\textbf{C}.
~\\\\
%\subsection*{Thermometer for 1D Bose gas}
\noindent
\uline{Thermometer for 1D Bose gas}

We calibrate the temperature of the 1D bosonic system accurately based on the one-body correlation function $g^{(1)}(z,z')=\langle \hat{\Psi}^\dagger(z')\hat{\Psi}(z)\rangle$, according to the method in Ref.~\cite{1d-thermometer}. Due to the strong quantum fluctuations in one dimension, the decay of  $g^{(1)}$ has a significant dependence on temperature, which allows us to use it as a thermometer. On the one hand, by performing the Fourier transform of the momentum distribution obtained through TOF measurement, we get integrated correlation function $G^{(1)}(z)=\int dz^{\prime} g^{(1)}(z^{\prime}+z, z^{\prime})$ for our 1D Bose gas. On the other hand, we can use the ab-initio quantum Monte Carlo (QMC) method to simulate the Lieb-Liniger Hamiltonian
%-------------------------------%
\begin{equation}\label{eq:Hamiltonian}
H_{LL} =\sum_{i} \Big[-\frac{\hbar^2}{2m}\frac{\partial^2}{\partial z_i^2}+V(z_i)\Big]+g_{\textrm{1D}}\sum_{i<j}\delta(z_i-z_j),
\end{equation}
%-------------------------------%
with $i$ and $j$ span the set of particles,
$g_{\textrm{1D}}$ the coupling constant and $V(z_j)$ the harmonic trap, similar to Refs.~\cite{yao-boseglass-2020,yao-crossoverD-2022}.
Our simulation is performed for a system with weighted averaged atom number $\bar{N}= \sum_{p,q} N_{p,q}^2/\sum_{p,q} N_{p,q}$ where $N_{p,q}$ is the atom number of the tube on site labelled by $p,q$. Thanks to the worm algorithm implementation~\cite{boninsegni-worm-short-2006}, we can compute $G^{(1)}(z)$ at different values of temperature which serves as a ruler. By comparing the experimental data with the QMC results, we can determine the temperature of the system within a resolution of $1$~nK. The numerical calculations make use of the ALPS scheduler library and statistical analysis tools~\cite{troyer1998,ALPS2007,ALPS2011}.

\section*{Supplementary Text}
%\subsection*{Lieb-Liniger model in a harmonic trap}
\noindent
\uline{Lieb-Liniger model in a harmonic trap}

Here we use the Lieb-Liniger Hamiltonian \cite{lieb1963exact} in a harmonic trap at zero temperature to describe the atoms in each 1D tube:
\begin{equation}
H_{LL}=\sum_i\{-\frac{\hbar^2}{2m}\frac{\partial^2}{\partial z^2_i}+V(z_i)\}+\sum_{i<j}g_{\textrm{1D}}\delta(z_i-z_j),
\end{equation}
where $V(z)=\frac{1}{2}m\omega^2 z^2$ is a harmonic potential along the $z$ direction with $\omega$ denoting the vibrational frequency. Using the Bethe ansatz and local density approximation, this model can be analytically solved and we can calculate the equilibrium state at zero temperature.
For each position $z$, we define the local dimensionless interaction strength $\gamma(z)={mg_{\textrm{1D}}}/{n_{\textrm{1D}}(z)\hbar^2}$ \cite{dunjko2001bosons} where $n_{\textrm{1D}}(z)$ is the local 1D density. Following the local density approximation, we define a cutoff radius $R_c$, where $n_{\textrm{1D}}(z)$ equals to zero for any $|z|>R_c$. Therefore, we define the average dimensionless interaction strength of one tube as
\begin{equation}
\gamma=\frac{\int_{-R_c}^{R_c} \textrm{d}z n_{\textrm{1D}}(z)\gamma(z)}{\int_{-R_c}^{R_c} \textrm{d}z n_{\textrm{1D}}(z)}.%=\frac{2mg_{\textrm{1D}}R_c}{N\hbar^2},
\end{equation}
%all the observables of equilibrium state are determined by the local dimensionless interaction strength $\gamma(z)={mg_{\textrm{1D}}}/{n_{\textrm{1D}}(z)\hbar^2}$ \cite{dunjko2001bosons}.   
In our experiment, we load atoms into about 3300 tubes of the 2D optical lattice, and each tube contains an ensemble of 1D Bose gas. Here the atom number in each tube is calculated by the density distribution of the initial BEC in the harmonic trap while the total atom number is $4\times10^4$. Therefore, the atom number $N_{p,q}$ can be calculated according to the Thomas-Fermi approximation as
\begin{equation}
    N_{p,q}=N_{0,0}\left[1-\frac{(pa)^2}{R_x^2}-\frac{(qa)^2}{R_y^2}\right],
\end{equation}
where $p$ and $q$ are integers and label the 2D lattice sites in the $x$-$y$ plane, $a=388.25~\textrm{nm}$ is the lattice constant, $N_{0,0}$ is the atom number in the central tube, and $R_x$ and $R_y$ are the Thomas-Fermi radii which can be calculated according to the trap geometries. 
%Here we experimentally measure the total atom number to determine $N(0,0)$. Consequently, the process provides us with $4\times10^4$ atoms in an array of 3300 1D tubes each with $N(i,j)$ atoms. The fitted distribution of $N(i,j)$ allows us to calculate the averaged atom number per tube $\bar{N}$ by 
Then we calculate the mean atom number per tube $\bar{N}$ by
\begin{equation}
    \bar{N}=\frac{\sum_{p,q}N_{p,q}^2}{\sum_{p,q}N_{p,q}},
\end{equation}
and the averaged interaction strength $\bar{\gamma}$ to characterize the 1D system
\begin{equation}
    \bar{\gamma}=\frac{\int^{\bar{R_c}}_{-\bar{R_c}}\textrm{d}z\bar{n}_{\textrm{1D}}(z)\bar{\gamma}(z)}{\bar{N}},
\end{equation}
where $\bar{n}_{\textrm{1D}}(z)$ is the density distribution of a tube with $\bar{N}$ atoms and $\bar{\gamma}(z)=mg_{\textrm{1D}}/\bar{n}_{\textrm{1D}}(z)\hbar^2$.

For each tube, there is a stretched-exponential decay of the atom number under its $\gamma$. To include the inhomogeneity of the atom number distribution theoretically in the theoretical simulations, we sum up each stretched-exponential decay according to its atom number and then fit it with an overall stretched-exponential function $N_0\exp\left[-(t/\tau_0)^\alpha\right]$. 
Under this method, the stretched exponent $\alpha$ is consistent with the predicted $\alpha$ by the averaged interaction strength $\bar{\gamma}$ under the Lieb-Liniger model. Therefore, 
in Fig.~2\textbf{C}, Fig.~3\textbf{C}, and Fig.~4\textbf{C}, 
we use the $\alpha$ obtained by this method to include the inhomogeneity of the atom number as the theoretical predicted value, and the
grey shades in the plots represent one standard deviations of fittings.
~\\\\
%\subsection*{Luttinger parameter $K$, anomalous dimension $\eta$, and interaction strength $\gamma$}
\noindent
\uline{Luttinger parameter $K$, anomalous dimension $\eta$, and interaction strength $\gamma$}

In this section, we will derive the relation between the Luttinger parameter $K$, the anomalous dimension $\eta$, and the interaction strength $\gamma$. 
The low-energy physics of the 1D gas can be described by the effective field theory,
%The low energy physics of 1D gas is investigated 
and the effective Hamiltonian \cite{giamarchi2003quantum} is
\begin{equation}
\hat{H}_{\textrm{eff}}=\int{\textrm{d}}z[\frac{\pi v_sK}{2\hbar}\hat{\Pi}^2+\frac{\hbar v_s}{2\pi K}(\partial_z\hat{\phi})^2],
\end{equation}
under the long-wavelength limit. Here $\hat\phi$ is the field operator and $\hat\Pi$ is the canonical momentum of $\hat\phi$. As bosonic operators, $\hat\phi$ and $\hat\Pi$ obey the bosonic commutation relation. The coefficient $K$ here is the Luttinger parameter and $v_s$ is the sound velocity. According to this Hamiltonian, the single-particle spectral function ${\mathcal A}_k(\omega)$ of 1D gas is proportional to  $(\omega^2-\Delta_k^2)^{-\eta}\Theta(\omega^2-\Delta_k^2)$ where $\Theta(x)$ is the heaviside step function and $\eta$ is the anomalous dimension, and $\eta$ is related to $K$ by $\eta=1-1/(4K)$. 

By applying Bethe ansatz solutions \cite{lieb1963exact,jiang2015understanding}, we can calculate the Luttinger parameter $K$. Based on the solutions, the energy of each particle at zero temperature is 
\begin{equation}
    \epsilon(n)=\frac{\hbar^2n^2}{2m}e(\gamma),\label{eq:zeroenergy}
\end{equation}
where $n$ is the mean density of 1D gas and $\gamma=mg_{\textrm{1D}}/n\hbar^2$ is the interaction strength, and the function $e(\gamma)$ is determined by
\begin{equation}
    e(\gamma)=\frac{\gamma^3}{\lambda^3(\gamma)}\int^1_{-1}g(z,\gamma)z^2{\textrm{d}}z.
\end{equation}
Here the functions $g(z,\gamma)$ and $\lambda(\gamma)$ are calculated through the Bethe ansatz solutions
\begin{eqnarray}
    g(z,\gamma)&=&\frac{1}{2\pi}\int^1_{-1}\frac{2\lambda(\gamma)}{\lambda^2(\gamma)+(y-z)^2}g(y,\gamma){\textrm{d}}y+\frac{1}{2\pi}, \\
    \lambda(\gamma)&=&\gamma\int_{-1}^1g(z,\gamma){\textrm{d}}z.\label{eq:zerolambda}
\end{eqnarray}
 In these equations, the only free parameter is the dimensionless interaction strength $\gamma$. Therefore, once it is determined, Eq.~\ref{eq:zeroenergy} to Eq.~\ref{eq:zerolambda} can be solved numerically. Based on these, the Luttinger parameter $K$ is obtained explicitly as a function of $\gamma$ as
\begin{equation}
    K=\frac{\pi}{\sqrt{3e(\gamma)-2\gamma\frac{{\textrm{d}}e(\gamma)}{{\textrm{d}}\gamma}+\frac{1}{2}\gamma^2\frac{{\textrm{d}^2}e(\gamma)}{{\textrm{d}}\gamma^2}}}.
\end{equation}

Now we move to the finite-temperature regime, and we want to prove that the Luttinger parameter $K$ is almost unchanged at finite temperatures.
We use the Yang-Yang equation to calculate this model at finite temperatures. The Luttinger parameter $K$ is calculated as $K=v_s/v_n$, where $v_s=\sqrt{ (\partial^2 F/\partial L^2)L^2/mn}$ is the sound velocity and $v_n=\partial^2 F/\partial N^2(2L/h)$ is the density stiffness. Here $n$ is the 1D density, $L$ is the system's length, $N$ is the total atom number, and $F$ is the free energy from the Yang-Yang equation. 
%Based on the numerical calculations, we find $K$ is almost unchanged under finite temperatures.

By introducing a dimensionless universal function $\tilde{f}(\gamma,\tilde{T})$ which is similar to $e(\gamma)$, we write out the total free energy as
%Therefore, At finite temperature we can also introduce a dimensionless universal function $\tilde{f}(\gamma,\tilde{T})$ (similar to $e(\gamma)$) as a function of $\gamma$ and dimensionless temperature $\tilde{T}=T/(n^2/2m)$, such that
\begin{eqnarray}
F=N\frac{\hbar^2n^2}{2m} \tilde{f}(\gamma,\tilde{T}),
\end{eqnarray}
where $\tilde{T}=2mk_B T/\hbar^2 n^2$ is the dimensionless temperature. 
It is intuitive to see that $\tilde{f}(\gamma,\tilde{T})$ becomes $e(\gamma)$ when $\tilde{T}$ approaches to zero. Therefore, we obtain the explicit formula of the finite-temperature Luttinger paramter $K$ as
\begin{eqnarray}
K=\!\frac{\pi}{\sqrt{3\tilde{f}\!-\!2\gamma\frac{\partial \tilde{f}}{\partial \gamma}\!+\!\frac{1}{2}\gamma^2\frac{\partial^2\tilde{f}}{\partial\gamma^2}\!+\!4\gamma\tilde{T}\frac{\partial^2 \tilde{f}}{\partial\gamma\partial\tilde{T}}\!+\!4\tilde{T}^2\frac{\partial^2\tilde{f}}{\partial\tilde{T}^2}}}.
\end{eqnarray}
In order to calculate $\tilde{f}(\gamma,\tilde{T})$, we need to solve the self-consistent Yang-Yang equations as follows,
\begin{eqnarray}
F&=&\frac{N}{n}\int_{-\infty}^{\infty}(\frac{\hbar^2k^2}{2m}-\varepsilon(k))\rho(k){\textrm{d}}k\nonumber\\
& &-\frac{Nk_BT}{n}\int_{-\infty}^{\infty}\!\!\!\!\!\!\! {\textrm{d}}kf(k)\log\left(1+e^{-\frac{\varepsilon(k)}{k_BT}}\right),\label{FE}\\
2\pi\rho(k)&=&\frac{1}{1+e^{\varepsilon(k)/k_BT}}\left(1+2g_{\textrm 1D}\int_{-\infty}^\infty {\textrm{d}}q\frac{\rho(q)}{\frac{m}{\hbar^2}g_{\textrm 1D}^2+\frac{\hbar^2(k-q)^2}{m}}\right),\\
\varepsilon(k)&=&-\mu+\frac{\hbar^2 k^2}{2m}-\frac{k_BTg_{\textrm 1D}}{\pi}\int_{-\infty}^{\infty}{\textrm{d}}q\frac{\log(1+e^{-\varepsilon(k)/k_BT})}{\frac{m}{\hbar^2}g_{\textrm 1D}^2+\frac{\hbar^2(k-q)^2}{m}},\\
\mu&=&\frac{1}{n}\int_{-\infty}^{\infty}(\frac{\hbar^2k^2}{2m}-\varepsilon(k))\rho(k){\textrm{d}}k\nonumber\\
&&+\frac{k_BT}{n}\int_{-\infty}^{\infty}{\textrm{d}}q \left(\frac{1}{2\pi}-f(q)\right)\log(1+e^{-\varepsilon(k)/k_BT}),\\
f(k)&=&\rho(k)(1+e^{\varepsilon(k)/k_BT})\label{YY}.
\end{eqnarray}
Combining Eq.~\ref{FE} to Eq.~\ref{YY}, we calculate $\tilde{f}(\gamma,\tilde{T})$ and the Luttinger parameter $K$ at finite temperature. In fig.~S1, we show that $K$ is almost unchanged under the finite-temperature effects.
\begin{figure}[h]
\begin{center}
\includegraphics[width=8cm]{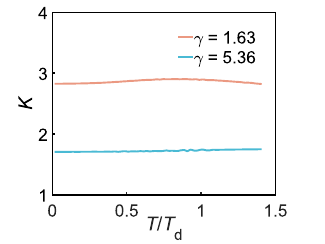}
\end{center}
\noindent {\bf Fig. S1.} \textbf{Dependence of Luttinger parameter on temperature}. The Luttinger parameter $K$ versus the rescaled temperature $T/T_d$, with $T_d = \hbar^2n^2/2mk_B$ denoting the degenerate temperature. Different colors represent different dimensionless interaction strengths $\gamma$.\\
\end{figure}
~\\\\
%\subsection*{Non-Hermitian linear response theory at finite temperature}
\noindent
\uline{Non-Hermitian linear response theory at finite temperature}

For a quantum system experiencing a dissipative perturbation, the dissipation term $ \hat{H}_{\textrm{diss}}$ can be written as a non-Hermitian Hamiltonian with Langevian noise operators in the form of
\begin{equation}
    \hat{H}_{\textrm{diss}}=-i\kappa \hat{\mathcal O}^\dagger\hat{\mathcal O}+\hat{\mathcal O}^\dagger\xi+\xi^\dagger\hat{\mathcal O}.
\end{equation}
Here $\xi$ and $\xi^\dagger$ are the Langevin noise operators, and $\kappa$ denotes the dissipation rate. Therefore, for any observable $\hat{\mathcal W}$, we consider the difference $\delta {\mathcal W}(t)$ at time $t$ compared with its value at the steady state without dissipation. The form of $\delta {\mathcal W}(t)$ is then calculated as 
\begin{equation}
\delta {\mathcal W}(t)=2\kappa\int {\textrm{d}}z\int_0^t {\textrm{d}}\tau\langle \hat{\mathcal O}^\dag(z,\tau)\hat{\mathcal W}(t)\hat{\mathcal O}(z,\tau) 
-\frac{1}{2}\{\hat{\mathcal O}(z,\tau)\hat{\mathcal O}(z,\tau),\hat{\mathcal W}(t)\}\rangle, \label{eq2}
\end{equation}
where $\{,\}$ corresponds to the anti-commutator.
To describe the one-body loss in our experiment, we use $\hat{\mathcal O}(z)=\int dk \hat{a}_k e^{ikz}$ where $\hat{a}_k$ is the bosonic annihilation operator at momentum $k$ mode, and we choose the atom number operator of momentum $k$ mode $\hat{n}_k=\hat{a}^\dagger_k \hat{a}_k$ as the observable $\hat{\mathcal W}$. Then we obtain
%In Lieb-Liniger regime, when the dissipative perturbation is set to be one-body loss with $\hat{\mathcal O}(x)=\hat{\psi}(x)=\int dk \hat{a}_k e^{ikx}$, where $\hat{a}_k$ is the bosonic annihilation operator at momentum $k$ mode, and we choose the atom number $\hat{\mathcal W}=\hat{n} as the observable, by reasonable approximation, Eq.~\ref{eq2} becomes
\begin{eqnarray}
\delta n_k(t)&=&-2\kappa n_k(0)\int_0^t {\textrm{d}}\tau g_k(\tau)g_k(-\tau), \label{eq3}\\
n_k(t)&=& n_k(0) e^{-{\mathcal F}_k(t)},\\
{\mathcal F}_k(t)&=&2\kappa\int_0^t {\textrm{d}}\tau g_k(\tau) g_k(-\tau),
\end{eqnarray}
here $g_k(t)$ is the single-particle correlation function formed as 
\begin{equation}
    g_k(t)=\langle [\hat{a}_k(t),\hat{a}^\dag_k(0)]\rangle=\int {\textrm{d}}\omega {\mathcal A}_k(\omega)e^{i\omega t}, \label{eqn:gA}
\end{equation}
where ${\mathcal A}_k(\omega)$ is the single-particle spectral function defined in the previous section. By integrating Eq. \ref{eqn:gA}, we find
\begin{eqnarray}
g_k(t)\approx c_0 |t|^{\eta-1} \cos(\Delta_k t),
\end{eqnarray}
where $c_0$ is a constant. 
%This approximation deviates from the accurate value of the integral only for very small t with $t\ll \Delta_k$. 
%For $t=0$, $g_k(0)=1$ as it is the normalization condition of the spectral function. One can see the dominant part of the integration 
Therefore, $\mathcal{F}_k(t)$ has a form as
\begin{eqnarray}
{\mathcal F}_k(t)&=&\kappa\frac{c_0^2}{2}\int_0^t {\textrm{d}}\tau\: \tau^{2\eta-2}\left[1+\cos(2\Delta_k \tau)\right]\nonumber\\
&=&(t/\tau_0)^{2\eta-1}+ G(2\Delta_k t),
\end{eqnarray}
where $\tau_0^{1-2\eta}=\kappa c_0^2/2$, and $G(2\Delta_k t)$ is a very small oscillation function whose period is $\pi/\Delta_k$. The major contribution term $(t/\tau_0)^{2\eta-1}$ in ${\mathcal F}_k(t)$ is the momentum-$k$ independent.
Therefore, this leads to $n_k(t)=n_k(0)\exp\left[-(t/\tau_0)^{2\eta-1}\right]$, and the total atom number $N(t)$ also follows the same form as,
\begin{equation}
N(t)=\sum_k n_k(t)= N(0)\exp\left[-(t/\tau_0)^{2\eta-1}\right]. \label{Eq:Nzero}
\end{equation}
The oscillation terms $G(2\Delta_k t)$ cancel each other because the ocillation frequencies $2\Delta_k$ are different. 
This is quite remarkable that the momentum average in presence of the trap indeed helps the emergence of universal behavior in $N(t)$.

For finite-temperature corrections, we need to consider the spectral function at non-zero temperature.
When the Luttinger liquid approximation is still valid in the Lieb-Liniger model, the single-particle correlation function is obtained by a conformal transformation which sends $t-z/v_s=z^*$ to $e^{i2\pi z^*/h\beta}$ and $\bar{z^*}=t+z/v_s$ to $e^{i2\pi \bar{z^*}/h\beta}$, and $\beta = 1/k_BT$ is the inverse temperature, where $k_B$ is the Boltzmann constant. Then the spatial correlation function becomes 
\begin{equation}
g(z,t;0,0)=\left|\left(\frac{\pi}{h\beta}\right)^2\frac{1}{\sinh\left(\pi z^*/h\beta\right)\sinh\left(\pi\bar{z^*}/h\beta\right)}\right|^{1-\eta}\Theta(t^2-z^2/v_s^2).
\end{equation}
Then the single-particle correlation function at finite temperature $g_k(t,T)$ is calculated as
\begin{eqnarray}
g_k(t,T)&=&\int_{-v_s|t|}^{v_s|t|} {\textrm{d}}z e^{ikz} g(z,t)\nonumber\\
&=&\int_{-v_s|t|}^{v_s|t|} {\textrm{d}}z \frac{ e^{ikz}\left(\frac{\pi}{h\beta}\right)^{2(\eta-1)}}{\left|\sinh^2\left(\frac{\pi t}{h\beta}\right)-\sinh^{2}\left(\frac{\pi z}{h\beta v_s}\right)\right|^{1-\eta}}\nonumber\\
&\approx&\int {\textrm{d}}z \frac{ e^{ikz}}{\left|\frac{h^2\beta^2}{\pi^2}\sinh^2\left(\frac{\pi t}{h\beta}\right)-\frac{z^2}{v_s^2}\right|^{1-\eta}}\nonumber\\
&=&g_k\left(\frac{h\beta}{\pi}\sinh\left(\frac{\pi t}{h\beta}\right), 0\right).
\end{eqnarray}
Because here we focus on the short-time correlation function, we have the approximation $\sinh(\pi z/h\beta v_s)h\beta/\pi\approx z/v_s$ in the integration region. 
In the last line of the above equation, $g_k(t,0)$ is obtained from the Fourier transformation from the space-time side. 
Therefore, the correlation function $g_k(t,T)$ at finite temperature $T$ can be approximately obtained by transformation $t\rightarrow \sinh(\pi t/h\beta)h\beta/\pi$ from the zero-temperature correlation function. Therefore, we have
\begin{eqnarray}
g_k(t)\approx c_0\left|\frac{h\beta}{\pi}\sinh\left(\frac{\pi t}{h\beta}\right)\right|^{-\eta}\!\!\!\!\!\cos\left[\Delta_k \frac{h\beta}{\pi}\sinh\left(\frac{\pi t}{h\beta}\right)\right].
\end{eqnarray}
Then ${\mathcal{F}}_k(t)$ is calculated as 
\begin{eqnarray}
{\mathcal F}_k(t)= \frac{1}{\tau_0^{2\eta-1}}\int_0^t {\textrm{d}}\tau\ \left(\sinh\frac{\pi \tau}{h\beta}\right)^{2\eta-2}\left(\frac{h\beta}{\pi}\right)^{2\eta-2}.
\end{eqnarray}
%where the oscillation terms are dropped because of the cancellation between different $k$.
Then $N(t)$ follows a modified decay form as
\begin{equation}
    N(t)=N(0)\exp\left[-\frac{1}{\tau_0^{2\eta-1}}\int_0^t {\textrm{d}}\tau\ \left(\sinh\frac{\pi \tau}{h\beta}\right)^{2\eta-2}\left(\frac{h\beta}{\pi}\right)^{2\eta-2}\right].\label{eq:exactN}
\end{equation}
Only when $t\ll h\beta/\pi$ is satisfied, $N(t)$ will have the same form as the result at zero temperature. Due to technical noises in the experiment, we have to choose $\pi t/h\beta\sim 1$. Therefore, we interrogate the time duration from 0 to $t_u$ to check the dissipative dynamics of $N(t)$. For $t\in[0,t_u]$, we fit $N(t)$ with the stretched-exponential function $N(0)\exp\left[-(t/\tau'_0)^{\alpha^*}\right]$, and we expect $\alpha^*$ becomes $2\eta-1$ when $t_u$ approaches to zero. In fig.~S2\textbf{A}, we first plot the exact $N(t)$ for both zero temperature (yellow curve) and finite one (cyan curve), then also plot the fitted function $N(0)\exp\left[-(t/\tau'_0)^{\alpha^*}\right]$ for comparison (red curve) where we set $t^*=\pi t_u/h\beta =4.8$ as the same one in the main text. It shows a good consistency between the finite-temperature curve $N(t)$ and the fitted one. In fig.~S2\textbf{B}, we plot $\alpha^*$ versus $t^*$ for different $\eta$ (or $\gamma$). 
%Here we can see $\int_0^{\frac{\pi t}{\beta}} d\tau (\sinh(\tau))^{-2(1-\eta)}=\int_0^{\frac{\pi t}{\beta}}\!\!\!\!d\tau \tau^{-2(1-\eta)}(\sinh\tau/\tau)^{-2(1-\eta)}\approx \int_0^{\frac{\pi t}{\beta}}\!\!\!\!d\tau \tau^{-2(1-\eta)}e^{-\frac{1-\eta}{3}\tau^2}$. When $t\pi/\beta\ll \sqrt{3/(1-\eta)}$, the integral is mainly $\int_0^{\frac{\pi t}{\beta}} \tau^{2(\eta-1)}$. Therefore typically there is a long time cutoff setting by temperature, which is $\sqrt{3/(1-\eta)}/\pi k_BT$.
% Your references go at the end of the main text, and before the
% figures.  For this document we've used BibTeX, the .bib file
% scibib.bib, and the .bst file Science.bst.  The package scicite.sty
% was included to format the reference numbers according to *Science*
% style.
\newpage
%\section*{Fig.S1}

\newpage
%\section*{Fig.S2}
\begin{figure}[h]
\begin{center}
\includegraphics[width=16cm]{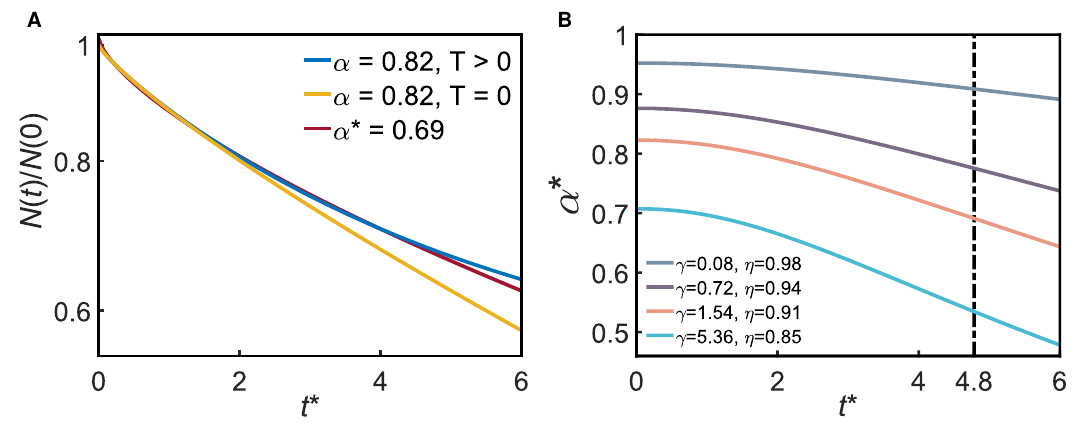}
\end{center}
\noindent {\bf Fig. S2.} \textbf{Finite-time correction}. (\textbf{A}) The cyan line is the dissipation curve in the form of Eq.~\ref{eq:exactN} with $\eta=0.91$. The yellow line is the dissipation curve with the same $\eta=0.91$ but at zero temperature formed in Eq.~\ref{Eq:Nzero}. Here, $t^*$ is the rescaled dissipation time $t^*=\pi k_B T t/h$. When we use the stretched-exponential function $N(0)\exp\left[-(t/\tau'_0)^{\alpha^*}\right]$ to fit the cyan line, the red line gives $\alpha^*=0.69$ when we set the time upper bound at $t^*=4.8$. Here the $y$ axis is in a logarithmic scale and the $x$ axis is in a linear scale. (\textbf{B}) For different $\gamma$ or $\eta$, as the $t^*$ increases, the fitted $\alpha^*$ decreases and the dashed line represents the $t^*$ we use in our experiment.
\end{figure}